%Paper: 9202009
%From: IMAMURA%JPNRIFP.BITNET@pucc.princeton.edu
%Date: Tue, 04 Feb 92 20:47:04 JST
%Date (revised): Wed, 05 Feb 92 17:10:44 JST

%%%%%%%%%%%%%%%%%%%%%%%%%%%%%%%%%%%
%%%%%                         %%%%%
%%%%   Preprint macro package  %%%%
%%%%%                         %%%%%
%%%%%%%%%%%%%%%%%%%%%%%%%%%%%%%%%%%

\magnification=\magstep1
\hsize=15.7truecm \vsize=23.4truecm
\baselineskip=6mm
\font\bg=cmbx10 scaled 1200
\footline={\hfill\ -- \folio\ -- \hfill}
\def\prenum#1{\rightline{#1}}
\def\date#1{\rightline{#1}}
\def\title#1{\centerline{\bg#1} \vskip 5mm}
\def\author#1{\centerline{#1} \vskip 3mm}
\def\address#1{\centerline{\sl#1}}
\def\abstract#1{{\centerline{\bg Abstract}} \vskip 3mm \par #1}

\def\references{{\centerline{\bg References}} \vskip 3mm}
\def\table{{\centerline{\bg Table Captions}} \vskip 5mm}
\def\chapter#1{\centerline{\bg#1} \vskip 5mm}
\def\section#1{\centerline{\bg#1} \vskip 5mm}
\def\endpage{\vfill \eject}
\def\lq{\char"5C}

\fontdimen5\textfont2=1.2pt
%%%%% Big zero in matrix (\bigzerou and \bigzerol) %%%%%
\font\bgg=cmr10 scaled\magstep4
\def\bigzerol{\smash{\hbox{\bgg 0}}}
\def\bigzerou{\smash{\lower1.7ex\hbox{\bgg 0}}}
%%%%% Semi-direct product %%%%%
\def\semiprod{ \ \mathrel \times \mathrel{\mkern-4mu} \mathrel {\vrule
height4.7pt width .2pt depth0pt \ } }
%%%%%%%%%%%%%%%%%%%%%%%%%%%%%%%%%%%%%%%%%%%%%%%%%%%%%%%%%%%%%%%

\prenum{KOBE--92--01}
\date{February 1992}

\vskip 10mm

\title{String Theories on the Asymmetric Orbifolds}
\centerline{\bg with Twist-Untwist Intertwining Currents \footnote{$^*$}{
\rm To appear in the proceedings of YITP Workshop on ``Recent Developments
in String and Field Theory'' at Kyoto, Japan on September 9-12, 1991.}}
\vskip 25mm

\author{Y. IMAMURA, M. SAKAMOTO, T. SASADA and M. TABUSE}
\address{Department of Physics, Kobe University}
\address{Nada, Kobe 657, Japan}
\vskip 3mm

\vfill

\abstract{We give various examples of asymmetric orbifold models to
possess intertwining currents which convert untwisted string states to
twisted ones, and vice versa, and see that such asymmetric orbifold models
are severely restricted. The existence of the intertwining currents leads
to the enhancement of symmetries in asymmetric orbifold models.}

\endpage

\section{1. Introduction}
String theory has been regarded as a candidate for the unified theory
including gravity and extensively been studied to construct
phenomenologically realistic models. Orbifold compactification [1] is one
of the most promising methods to build them and the search for realistic
orbifold models has been continued by many groups and various models have
been proposed [2-5]. However, we have not yet found any satisfactory
orbifold models to describe our real world. Therefore, it would be of
great importance to classify all the compactifications of string theories
on orbifolds thoroughly. In this paper, we shall investigate symmetries
between untwisted and twisted string states on asymmetric orbifolds [6]
\footnote{$^{\ddag 1}$}{Some examples have been discussed in refs.[1,7,8]
and in our previous paper [9].}.

Suppose that there exists an intertwining current operator\footnote{$^{
\ddag 2}$}{This current is a twisted state emission vertex operator
[7,8,10-12] with the conformal weight (1,0) or (0,1), whose explicit
construction is not easy in general.} which converts string states in the
untwisted sector to string states in the twisted sector in an asymmetric
orbifold model. This current operator will correspond to a state of the
conformal weight (1,0) (or (0,1)) in the twisted sector and connect the
ground state of the untwisted sector to the (1,0) (or (0,1)) twisted
state. Therefore, the existence of a (1,0) (or (0,1)) twisted state
implies the appearance of a symmetry between the untwisted and twisted
sectors. It leads to the enhancement of symmetries in asymmetric orbifold
models.
It should be emphasized that it does not occur in the case of symmetric
orbifolds because the left- and right-conformal weights, $h$ and $\bar h$,
of a ground state of any twisted sector are both positive for symmetric
orbifolds and hence no (1,0) or (0,1) state appears in any twisted sector.

In the next section, we briefly review general properties of asymmetric
orbifolds and give the modular invariance conditions of the one-loop
partition function. In section 3, we investigate ${\bf Z}_N$-automorphisms
of the lattice defining the orbifolds.  In section 4, we prove the
``torus-orbifold equivalence'' [1,6,13-15]. This equivalence is used to
determine the symmetries of orbifold models. In section 5, we give various
examples of asymmetric ${\bf Z}_2$-, ${\bf Z}_3$- and  ${\bf
Z}_N$-orbifold models which possess (1,0) twisted states and show the
enhancement of symmetries in the asymmetric orbifold models. Finally in
section 6, we present our conclusion.

%\endpage
\vskip 10mm

\section{2. Asymmetric ${\bf Z}_N$-Orbifolds}

In the construction of toroidal orbifolds, we start with a
$D$-dimensional toroidally compactified closed bosonic string theory which
is specified by a $D+D$-dimensional Lorentzian even self-dual lattice $
\Gamma^{D,D}$ [16]. The left- and right-moving momentum $(p_L^I, p_R^I)$ $
(I=1,\cdots , D)$ lies on the lattice $\Gamma^{D,D}$. For simplicity, we
will restrict our considerations to asymmetric ${\bf Z}_N$-orbifolds and
choose $\Gamma^{D,D}$ to be of the form
$$ \Gamma^{D,D}=\{\ (p_L^I, p_R^I)\  | \  p_L^I-p_R^I \in \Lambda {\quad
\rm and \quad} p_L^I, p_R^I \in \Lambda^*\}, \eqno{(2.1)}$$
where $\Lambda$ is a $D$-dimensional Euclidean lattice and $\Lambda^*$ is
the dual lattice of $\Lambda$. Note that $\Gamma^{D,D}$ is even self-dual
if $\Lambda$ is even integral. Let $g$ be a group element which generates
the asymmetric ${\bf Z}_N$-transformation. The $g$ is defined by
$$ g : \quad (X_L^I, X_R^I) \rightarrow (U^{IJ}X_L^J, X_R^I), \qquad I,J
=1,...,D, \eqno{(2.2)}$$
where $X_L^I$, $(X_R^I)$ is a left- (right-) moving string coordinate and
$U$ is a rotation matrix which satisfies $U^N={\bf 1}$. The ${\bf
Z}_N$-transformation has to be an automorphism of the lattice $
\Gamma^{D,D}$, i.e.,
$$ (U^{IJ}p_L^J, p_R^I) \in \Gamma^{D,D} {\qquad \rm for \  all \  }
(p_L^I, p_R^I) \in \Gamma^{D,D}. \eqno{(2.3)}$$

In the untwisted sector, the boundary condition of the string coordinate
is the same as the torus case, so that the left- and right-moving string
coordinate will be expanded as
$$ X^I_L(z)=x^I_L-ip^I_L ln z +i\sum_{n \ne 0}{1\over n}\alpha^I_{L
n}z^{-n},$$
$$ X^I_R(\bar z)=x^I_R-ip^I_R ln {\bar z} +i\sum_{n \ne 0}{1\over n}
\alpha^I_{R n}{\bar z}^{-n}.\eqno{(2.4)}$$
Now we will introduce the operator $R_{(0)}$, which induces the ${\bf
Z}_N$-transformation, i.e.,
$$ R_{(0)}(X^I_L(z),X^I_R(\bar z))R_{(0)}^{-1}=(U^{IJ}X^J_L(z),X^I_R(\bar
z)).\eqno{(2.5)}$$
To explicitly construct $R_{(0)}$ in terms of the operators of the mode
expansion of the string coordinate, it will be convenient to use a complex
coordinate. Since $U$ is an orthogonal matrix, it can be diagonalized by a
unitary matrix $M$:
$$ MUM^\dagger = U_{diag}.\eqno{(2.6)}$$
Since $U^N=1$, we may write
$$ U_{diag}=\pmatrix{\omega^{r_1} &              &         &  \bigzerou
\cr
                                  & \omega^{r_2} &         &
\cr
                                  &              & \ddots  &
\cr
                     \bigzerol    &              &         & \omega^{r_D}
\cr
                    }, \eqno{(2.7)}$$
where $0 \le r_I \le N-1 $ $(r_I \in {\bf Z})$ and $\omega = e^{2\pi i /
N}$. The set of eigenvalues $\{ \omega^{r_I} \}$ is identical to the set
of $\{ \omega^{-r_I} \}$ because $U$ is an orthogonal matrix. Thus, we may
write the eigenvalues of $U$ as\footnote{$^{\ddag 3}$}{Here we assumed
that the dimension $D$ is even integer. In fact, $D$ is even for all the
models we consider.}
$$\{ \omega^{r_I} \ {\rm and} \ \omega^{-r_I}, \quad I=1,2,\cdots,{D\over
2} \}.$$
In terms of $\gamma^I_{L n} \equiv M^{IJ}\alpha^J_{L n}$ $(n \in {\bf Z}
>0)$, the operator $R_{(0)}$ is given by
$$ R_{(0)}= \omega^{-\sum^D_{I=1}\sum^\infty_{n=1}{r_I\over n}{\gamma^I_{L
n}}^\dagger \gamma^I_{L n} } \sum_{(p^I_L,p^I_R)\in \Gamma^{D,D}} |p^I_L,
p^I_R><U^{IJ}p^J_L, p^I_R|.\eqno{(2.8)}$$

The one-loop partition function in the untwisted sector is given by
$$ Z^{(0)}(\tau)={1\over N}\sum^{N-1}_{m=0} Z(1,U^m;\tau),\eqno{(2.9)}$$
where
$$ Z(1,U^m;\tau)={\rm Tr}[(R_{(0)})^m q^{L_0-{D\over 24}}{\bar q}^{{\bar
L}_0-{D\over 24}}],$$
$$ q=e^{i2\pi\tau},$$
$$ L_0=\sum_{I=1}^D\{{1\over 2}(p^I_L)^2+\sum_{n=1}^\infty\alpha^I_{L -n}
\alpha^I_{L n}\}, $$
$$ {\bar L}_0=\sum_{I=1}^D\{{1\over 2}(p^I_R)^2+\sum_{n=1}^\infty
\alpha^I_{R -n}\alpha^I_{R n}\}. $$
Calculating the trace of each terms, we have
$$ Z(1,1;\tau)={1\over |\eta(\tau)|^{2D}}\sum_{(p^I_L,p^I_R)\in
\Gamma^{D,D}}q^{{1\over 2}(p^I_L)^2}{\bar q}^{{1\over 2}(p^I_R)^2}, \eqno
(2.10)$$
$$ Z(1,U^m;\tau)={1\over |\eta(\tau)|^{2D}}\prod_{I=1}^{D/2}[{-2sin(
\pi{[mr_I]\over N})(\eta(\tau))^3\over\vartheta_1({[mr_I]\over N}|\tau)}]
\sum_{(p^I_L,p^I_R)\in\Gamma^{D,D}_{inv}}q^{{1\over 2}(p^I_L)^2}{\bar
q}^{{1\over 2}(p^I_R)^2}, \eqno(2.11)$$
$$ \qquad\qquad{\rm for}\quad m=1,2,...,N-1,$$
where $\eta(\tau)$ is the Dedekind $\eta$-function and $\vartheta_1(\nu|
\tau)$ is the Jacobi theta function:
$$\eta(\tau)=q^{1/24}\prod_{n=1}^\infty(1-q^n),$$
$$\vartheta_1(\nu|\tau)=\sum_{n=-\infty}^\infty exp\{i\pi(n+{1\over2})^2
\tau+i2\pi(n+{1\over2})(\nu+{1\over2})\},$$
$[ x ]$ denotes that $[x]=x$ mod $N$ and $0\le[x]<N$ for any integer $x$
and
$$ \Gamma^{D,D}_{inv}=\{(p^I_L,p^I_R)\in \Gamma^{D,D}|(p^I_L,p^I_R)=(
(U^m)^{IJ}p^J_L,p^I_R)\}.\eqno(2.12)$$
Note that modular invariance of $Z(1,1;\tau)$ requires that the lattice $
\Gamma^{D,D}$ must be a $(D+D)$-dimensional Lorentzian even self-dual
lattice.

In the $U^\ell$-twisted sector ($\ell=1,2,...,N-1)$, the string
coordinate will obey the following $U^\ell$-twisted boundary condition:
$$ X^I_L(e^{2\pi i}z) = (U^\ell)^{IJ}X^J_L(z) + (shift), $$
$$ X^I_R(e^{-2\pi i}\bar z) = X^I_R(\bar z) + (shift). \eqno(2.13)$$
Thus $X^I_L(z)$ and $X^I_R(\bar z)$ will be expanded as\footnote{$^{\ddag
4}$}{In this paper, we will consider ${\bf Z}_N$-orbifold models, in which
${\bf Z}_N$-transformation leaves only the origin fixed. Therefore $U^
\ell$ has no eigenvalue of one.}
$$ X^I_L(z)=x^I_L+i\sum_{n_J \in Z+{[\ell r_J]\over N} > 0}{1\over n_J}\{
(M^\dagger)^{IJ}\gamma^J_{L n_J}z^{-n_J}-M^{JI}{\gamma^J_{L n_J}}^\dagger
z^{n_J}\},$$
$$ X^I_R(\bar z)=x^I_R-ip^I_R ln {\bar z} +i\sum_{n \ne 0}{1\over n}
\alpha^I_{R n}{\bar z}^{-n}.\eqno(2.14)$$
As in the untwisted sector, the partition function of the $U^\ell$-twisted
sector consists of $N$ parts:
$$ Z^{(\ell)}(\tau)={1\over N}\sum^{N-1}_{m=0} Z(U^\ell,U^m;\tau),\qquad
\ell=1,2,...,N-1,\eqno(2.15)$$
where
$$ Z(U^\ell,U^m;\tau)={\rm Tr}[(R_{(\ell)})^m q^{L_0-{D\over 24}} {\bar
q}^{{\bar L}_0-{D\over 24}}]$$
and $R_{(\ell)}$ is the ${\bf Z}_N$-transformation operator in the $U^
\ell$-twisted sector:
$$R_{(\ell)}(X^I_L(z),X^I_R(\bar z))R_{(\ell)}^{-1}=(U^{IJ}X^J_L(z),X^I_R(
\bar z)).$$
However the twisted Hilbert space is not obvious in the case of asymmetric
orbifolds. In the following, to understand this Hilbert space, we will use
the modular transformation properties of the one-loop partition function
as a guiding principle.
In order that the partition function is modular invariant, each term in
the partition function should transform as
$$ Z(U^\ell,U^m;\tau+1)=Z(U^\ell,U^{m+\ell};\tau), $$
$$ Z(U^\ell,U^m;-1/\tau)=Z(U^{-m},U^{\ell};\tau). \eqno(2.16)$$
Therefore, we define $Z(U^\ell,1;\tau)$ as follows:
$$\eqalign{&Z(U^\ell,1;\tau)\equiv Z(1,U^{-\ell};-1/\tau)\cr
 &={{\sqrt{det(1-U^\ell)}\over V_{\Gamma^{D,D}_{inv}}}{q^{h_\ell}\over |
\eta(\tau)|^{2D}}\prod_{I=1}^{D/2}[{(\eta(\tau))^3\over i q^{[\ell r_I]
\over 2N}\vartheta_1({[\ell r_I]\over N}\tau|\tau)}}] \sum_{(p^I_L,p^I_R)
\in {\Gamma^{D,D}_{inv}}^*}q^{{1\over 2}(p^I_L)^2}{\bar q}^{{1\over 2}
(p^I_R)^2},\cr} \eqno(2.17)$$
where $V_{\Gamma^{D,D}_{inv}}$ is the volume of the unit cell of $
\Gamma^{D,D}_{inv}$, $h_\ell$ is the conformal weight of $U^\ell$-twisted
vacuum:
$$ h_\ell={1\over 4}\sum_{I=1}^D{[\ell r_I]\over N}(1-{[\ell r_I]\over N})
\eqno(2.18)$$
and ${\Gamma^{D,D}_{inv}}^*$ is the dual lattice of $\Gamma^{D,D}_{inv}$.

The remaining parts of the partition function are defined as follows
\footnote{$^{\ddag 5}$}{If $(N,\ell)\ne 1$, undetermined parts still
remain, where $(N,\ell)$ denotes the greatest common divisor of $N$ and $
\ell$. These are defined by other partition functions $Z(U^\ell,U^m;\tau)$
using the modular transformation:$\tau \rightarrow -1/\tau$.}:
$$\eqalign{&Z(U^\ell,U^{\ell m};\tau)\equiv Z(U^\ell,1;\tau+m)\cr
  &={{\sqrt{det(1-U^\ell)}\over V_{\Gamma^{D,D}_{inv}}} {e^{2\pi i m h_
\ell}q^{h_\ell}\over |\eta(\tau)|^{2D}}\prod_{I=1}^{D/2}[{(\eta(\tau))^3
\over i e^{i \pi m {[\ell r_I]\over N}} q^{[\ell r_I]\over 2N}\vartheta_1
({[\ell r_I]\over N}(\tau+m)|\tau)}}]\cr
  &\qquad\qquad\qquad\times \sum_{(p^I_L,p^I_R)\in {\Gamma^{D,D}_{inv}}^*}
e^{i\pi m ((p^I_L)^2-(p^I_R)^2)} q^{{1\over 2}(p^I_L)^2}{\bar q}^{{1\over
2}(p^I_R)^2}.\cr} \eqno(2.19)$$

As stated above, we have obtained $Z(U^\ell,U^{\ell m};\tau)$ $(\ell\ne
0)$ from $Z(1,U^{-\ell};\tau)$ using the modular transformation
properties. Considering these partition functions, we now find that $Z(U^
\ell,U^{\ell m};\tau)$ $(\ell\ne 0)$ given in eq.(2.19) can also be
obtained from the trace formula in the operator formalism:
$$ Z(U^\ell,U^{\ell m};\tau)={\rm Tr}[({R_{(\ell)}})^{\ell m} q^{L_0-{D
\over 24}}{\bar q}^{{\bar L}_0-{D\over 24}}],\eqno(2.20)$$
where
$$ L_0=\sum_{I=1}^D \sum_{n_I\in Z+ {[\ell r_I]\over N} >0} {
\gamma^I_{Ln_I}}^\dagger \gamma^I_{Ln_I} + h_\ell, $$
$$ {\bar L}_0=\sum_{I=1}^D\{{1\over 2}(p^I_R)^2+\sum_{n=1}^\infty
\alpha^I_{R -n}\alpha^I_{R n}\}, $$
$$ (R_{(\ell)})^\ell=e^{2\pi i(L_0-{\bar L}_0)}.$$
Note that the trace of momentum $(p^I_L,p^I_R)$ is over ${
\Gamma^{D,D}_{inv}}^*$ and the number of degeneracy of the ground states
in the $U^\ell$-twisted sector is given by
$$ n={\sqrt{det(1-U^\ell)} \over V_{\Gamma^{D,D}_{inv}}}.\eqno(2.21)$$

Suppose that $(N,\ell)=d$, where $(N,\ell)$ denotes the greatest common
divisor of $N$ and $\ell$. Since $Z(U^\ell,U^m;\tau)$ has to be invariant
under the modular transformation $\tau \rightarrow \tau + N/d$ because of
$U^{N}={\bf 1}$, the necessary condition for the modular invariance is
$$ {N\over d} (L_0 - {\bar L}_0 ) = 0 \ {\rm mod}\  1. \eqno(2.22)$$
This is called the left-right level matching condition and it has been
proved that this condition is also a sufficient condition for modular
invariance [6,17]. This condition can equivalently be rewritten as
follows:
$$ {N\over d} h_\ell = 0 \ {\rm mod}\  1,\eqno(2.23)$$
$$ {N\over d} ( (p^I_L)^2 - (p^I_R)^2 ) = 0 \ {\rm mod}\ 2  \qquad  {\rm
for \  all}\  (p^I_L,p^I_R) \in {\Gamma^{D,D}_{inv}}^* .\eqno(2.24)$$

%\endpage
\vskip 10mm

\section{3. Automorphism of $\Gamma^{D,D}$}
In the following we consider asymmetric ${\bf Z}_N$-orbifolds, where $
\Lambda$ in eq.(2.1) is a root lattice of a simply-laced Lie group $G$
(i.e., $\Lambda=\Lambda_R(G)$ ) and the squared length of the root is
normalized to two. For simplicity, we will restrict our considerations to
the case that ${\bf Z}_N$-transformation leaves only the origin fixed. We
will first investigate the automorphisms of $\Gamma^{D,D}$ ($D=$rank$G$).
The group of ``asymmetric'' automorphisms of $\Gamma^{D,D}$, Aut$
\Gamma^{D,D}$, is defined by
$$ (U^{IJ}p^I_L, p^I_R) \in \Gamma^{D,D} \qquad {\rm for \ all}\quad
(p^I_L, p^I_R) \in \Gamma^{D,D}, \eqno(3.1)$$
where $U \in {\rm Aut} \Gamma^{D,D}$. This means that Aut$\Gamma^{D,D}$
must be contained in the group of automorphisms of the root lattice $
\Lambda_R(G)$, Aut$\Lambda_R(G)$.

Aut$\Lambda_R(G)$ is semi-direct product of two groups [18]:
$$ {\rm Aut}\Lambda_R(G) = W_G \semiprod \Gamma_G,\eqno(3.2)$$
where $W_G$ is the Weyl group of the root system of $G$, i.e., the group
generated by the Weyl reflection of simple roots, and $\Gamma_G$ is
\footnote{$^{\ddag 6}$}{See ref.[18] for detail.}
$$ \Gamma_G=\{ \sigma\in {\rm Aut} \Lambda_R(G) | \sigma(\Delta)=\Delta\}.
\eqno(3.3)$$
Here $\Delta$ is a fixed basis of $\Lambda_R(G)$, i.e., $\Delta=\{
\alpha_1,\alpha_2,\cdots,\alpha_D\}$ and $\alpha_i$ $(i=1,2,\cdots,D)$ is
a simple root of $G$. $\Gamma_G$ corresponds to the group of symmetries of
the Dynkin diagram of $G$. Any element of $W_G$ transforms $p^I_L$ ($\in{
\Lambda_R(G)}^*=\Lambda_W(G)$) to $U^{IJ}p^J_L$ in the same conjugacy
class of $G$ but an element in $\Gamma_G$ does not. Therefore from eq.
(3.1) Aut$\Gamma^{D,D}$ must be in $W_G$ not in full Aut$\Lambda_R(G)$.

Every element in the Weyl group, $w\in W_G$, can be written as a product
of rank$G$ or less Weyl reflections of linearly independent roots [19,20],
i.e.,
$$ w=w_1w_2...w_k, \qquad 1\le k\le {\rm rank} G,\eqno(3.4)$$
where $w_i$ denotes a Weyl reflection with respect to a root $\alpha_i$
which needs not to be a simple root. Reflection of $k$ $(\le {\rm rank}
G)$ linearly independent roots in a rank$G$-dimensional vector space
leaves a (rank$G-k$)-dimensional subspace fixed. Hence Weyl elements to
leave only the origin fixed must be reflections of $k=$rank$G$ linearly
independent roots. These elements are given in ref.[19] and are summarized
as follows:

In the case of $G=SU(\ell+1)$, a Weyl element leaves only the origin
fixed only if the order of the element is $\ell+1$ and is prime. This
element is given by
$$ w=w_1w_2...w_\ell,\eqno(3.5)$$
where $w_i$ is the Weyl reflection of a simple root $\alpha_i$. Under this
element the simple root transforms as follows:
$$ \alpha_i \rightarrow \alpha_{i+1}, \qquad i=1,2,\cdots , \ell,$$
and
$$ \alpha_{\ell+1}\equiv-(\alpha_1+\alpha_2+...+\alpha_\ell) \rightarrow
\alpha_1.$$

In the case of $G=SO(2\ell$), the order of allowed Weyl elements is
$$ 2^N, 2^{N-1}, ..., 2 \qquad {\rm for} \  \ell=2^N p \quad (p={\rm odd \
integer}). $$
The root vectors of $SO(2\ell)$ will be given by $\pm e_a \pm e_b$ $(a\ne
b)$, where $e_a$ $(a=1,2,\cdots,\ell)$ is an orthogonal unit vector. A
cyclic transformation of $e_{k_1},e_{k_2},...,e_{k_m}$ such that
$$ e_{k_1} \rightarrow e_{k_2} \rightarrow ... \rightarrow e_{k_m}
\rightarrow -e_{k_1} \rightarrow -e_{k_2} \rightarrow ... \rightarrow
-e_{k_m} \rightarrow e_{k_1} $$
is denoted by $[{\overline m}]$, then the transformation of the order
$2^i$ $(i=1,...,N)$ is expressed by
$$ 2^i : \quad [\underbrace{{\overline {2^{i-1}}},{\overline
{2^{i-1}}},...,{\overline {2^{i-1}}}}_{ 2^{N+1-i}p \ {\rm times}}]. $$
For example, the transformations of order 2 and $2^2$ are given by
$$e_1\rightarrow -e_1\rightarrow e_1,\quad e_2\rightarrow -e_2\rightarrow
e_2,\quad\cdots,\quad e_\ell\rightarrow -e_\ell\rightarrow e_\ell$$
and
$$\eqalign{
&e_1\rightarrow e_2\rightarrow -e_1\rightarrow -e_2\rightarrow e_1,\cr
&e_3\rightarrow e_4\rightarrow -e_3\rightarrow -e_4\rightarrow e_3,\cr
&\qquad\qquad\qquad\cdots\cr
&e_{\ell-1}\rightarrow e_\ell\rightarrow -e_{\ell-1}\rightarrow -e_\ell
\rightarrow e_{\ell-1},\cr}$$
respectively.

In the case of $G=E_6,E_7,E_8$, the order of allowed Weyl elements is
$$ \eqalign{E_6 &: \quad 3,9, \cr
            E_7 &: \quad 2,   \cr
            E_8 &: \quad 2,3,4,5,6,8,10,12,15,20,24,30.\cr} $$

%\endpage
\vskip 10mm

\section{4. Torus-Orbifold Equivalence}
In this paper we shall investigate the symmetries of asymmetric ${\bf
Z}_N$-orbifold models with intertwining currents. However the explicit
construction of these currents is not easy in general. Therefore we may
rewrite the orbifold model into an equivalent torus model using the
``torus-orbifold equivalence'' [1,6,13-15] and investigate symmetries of
this torus model instead of the orbifold model. The ``torus-orbifold
equivalence'' tells us that {\sl any compactified closed bosonic string
theory on a ${\bf Z}_N$-orbifold is equivalent to that on a torus if the
rank of the gauge symmetry of strings on the orbifold is equal to the
dimension of the orbifold}. It may be instructive to give a proof of the
``torus-orbifold equivalence'' here. The proof will follow ref.[15].

Let us consider a $D$-dimensional torus model associated with the root
lattice $\Lambda_R(G)$ of a simply-laced Lie group $G$ ($D={\rm rank}G$).
Then this model has the affine Kac-Moody algebra $\hat g + \hat g$, which
can be constructed in the vertex operator representation {\sl \` a la}
Frenkel and Kac [21]:
$$\eqalign{ P^I_L(z)&\equiv i\partial_z X^I_L(z),\cr
     V_L(\alpha ; z)&\equiv : exp\{i\alpha \cdot X_L(z)\} : \cr} \eqno
(4.1)$$
and
$$\eqalign{ P^I_R(\bar z)&\equiv i\partial_{\bar z} X^I_R(\bar z),\cr
     V_R(\alpha ; \bar z)&\equiv : exp\{i\alpha \cdot X_R(\bar z)\} : .
\cr}\eqno(4.2)$$
A ${\bf Z}_N$-orbifold model is obtained by modding out of this torus
model by a ${\bf Z}_N$-rotation which is an automorphism of the lattice
defining the torus. Since every physical string state on the ${\bf
Z}_N$-orbifold is invariant under the ${\bf Z}_N$-transformation, the
gauge symmetry $G_0$, which is the invariant subalgebra of $G$ under ${\bf
Z}_N$-transformation, will appear in the spectrum.

In the case of ${\rm rank}G_0=D$, the ${\bf Z}_N$-invariant operator
$P'^I_L(z)$, $P'^I_R(\bar z)$ $(I=1,2,\cdots ,D)$ can be constructed from
suitable linear combinations of $P^I_L(z)$, $V_L(\alpha ;z)$ and $P^I_R(
\bar z)$, $V_R(\alpha ;\bar z)$ such that
$$ R_{(\ell)}(P'^I_L(z), P'^I_R(\bar z))R_{(\ell)}^{-1} = (P'^I_L(z),
P'^I_R(\bar z))\eqno(4.3)$$
and
$$ P'^I_L(w)P'^J_L(z)={\delta^{IJ}\over (w-z)^2}+\cdots, $$
$$ P'^I_R(\bar w)P'^J_R(\bar z)={\delta^{IJ}\over (\bar w-\bar z)^2}+
\cdots,\eqno(4.4) $$
where the operator $R_{(\ell)}$ is the ${\bf Z}_N$-transformation operator
in the $U^\ell$-sector ($\ell=0$ for the untwisted sector and $\ell=1,2,
\cdots,N-1$ for the $U^\ell$-twisted sector). It follows from (4.4) that
$P'^I_L(z)$, $P'^I_R(\bar z)$ can be expanded as
$$ P'^I_L(z) \equiv i\partial_z X'^I_L(z) \equiv \sum_{n\in Z}\alpha'^I_{L
n}z^{-n-1},$$
$$ P'^I_R(\bar z) \equiv i\partial_{\bar z} X'^I_R(\bar z) \equiv \sum_{n
\in Z}\alpha'^I_{R n}{\bar z}^{-n-1}\eqno(4.5)$$
with
$$[\alpha'^I_{Lm},\alpha'^J_{Ln}]=m\delta^{IJ}\delta_{m+n,0},$$
$$[\alpha'^I_{Rm},\alpha'^J_{Rn}]=m\delta^{IJ}\delta_{m+n,0}.$$

In this basis the vertex operator will be written as
$$ V'(k_L, k_R ; z)=:exp\{ik_L\cdot X'_L(z)+ik_R\cdot X'_R(\bar z)\}:,
\qquad (k_L, k_R) \in \Gamma_{G}^{D,D}.\eqno(4.6)$$
Since the operators $P'^I_L(z)$, $P'^I_R(\bar z)$ are invariant under the
${\bf Z}_N$-transformation, the vertex operator will transform as
$$ R_{(\ell)}V'(k_L,k_R,z)R_{(\ell)}^{-1} = e^{i2\pi(k_L \cdot v_L-k_R
\cdot v_R)}V'(k_L,k_R,z), \eqno(4.7)$$
where $(v_L, v_R)$ is some constant vector. Therefore, $R_{(\ell)}$ will
be given by
$$ R_{(\ell)}=\eta_{(\ell)}exp\{i2\pi(\hat p'_L\cdot v_L - \hat p'_R\cdot
v_R)\},\eqno(4.8)$$
where $\eta_{(\ell)}$ is a constant phase. Thus, the string coordinate in
the new basis transforms as
$$ (R_{(\ell)})^{\ell}(X'^I_L(z), X'^I_R(\bar z))(R_{(\ell)}^{-1})^{\ell}
= (X'^I_L(z)+2\pi \ell v^I_L, X'^I_R(\bar z)-2\pi \ell v^I_R).\eqno(4.9)$$
This implies that the string coordinate $(X'^I_L(z), X'^I_R(\bar z))$
obeys the boundary condition
$$ (X'^I_L(e^{2\pi i}z), X'^I_R(e^{-2\pi i}\bar z)) = (X'^I_L(z)+2\pi \ell
v^I_L, X'^I_R(\bar z)-2\pi \ell v^I_R) + (torus \ \ shift),\eqno(4.10)$$
and hence that
$$ (p'^I_L, p'^I_R) \in \Gamma_{G}^{D,D} + \ell (v^I_L, v^I_R).\eqno
(4.11)$$
In the new basis, $(R_{(\ell)})^{\ell}$ will be given by
$$(R_{(\ell)})^{\ell}=e^{i2\pi(L'_0-{\bar L}'_0)}\eqno(4.12)$$
because of the relation $(2.16)$. Here $L'_0$and ${\bar L}'_0$ are
$$ L'_0=\sum_{I=1}^D\{{1\over 2}(p'^I_L)^2+\sum_{n=1}^\infty\alpha'^I_{L
-n}\alpha'^I_{L n}\}, $$
$$ {\bar L}'_0=\sum_{I=1}^D\{{1\over 2}(p'^I_R)^2+\sum_{n=1}^\infty
\alpha'^I_{R -n}\alpha'^I_{R n}\}. \eqno(4.13)$$
Then it follows from eqs.(4.8),(4.11),(4.12) and (4.13) that $\eta_{(
\ell)}$ is given by
$$ \eta_{(\ell)}=exp\{-i\pi\ell((v^I_L)^2-(v^I_R)^2)\}.$$
Since $(v^I_L, v^I_R)$ will correspond to one of momentum eigenvalues of
the ground state in the $U$-twisted sector[15] and  $(R_{(\ell)})^N = 1$,
$(v^I_L, v^I_R)$ must satisfy
$$\eqalign{ {1\over 2}(v^I_L)^2&=h_1,\cr
            {1\over 2}(v^I_R)^2&={\bar h}_1,\cr
             N(v^I_L, v^I_R) &\in \Gamma_{G}^{D,D},\quad(m(v^I_L, v^I_R)
\notin \Gamma_{G}^{D,D}, m=1,2,\cdots,N-1),}\eqno(4.14)$$
where $h_1$ ($\bar h_1$) is the conformal weight of the ground state of
left- (right-) mover in $U$-twisted sector.

Every physical state in the $U^\ell$-sector must obey the condition $R_{(
\ell)}=1$ because it must be invariant under the ${\bf
Z}_N$-transformation. Thus the allowed momentum eigenvalues $(p'^I_L,
p'^I_R)$ of the physical state in the $U^\ell$-sector are restricted to
$$ (p'^I_L, p'^I_R) \in \Gamma_{G}^{D,D} + \ell (v^I_L, v^I_R) \ \  {\rm
with }\ \  p'_L\cdot v_L-p'_R\cdot v_R -{1\over 2} \ell ((v^I_L)^2-
(v^I_R)^2) \in {\bf Z}$$
$$ \qquad\qquad{\rm for \ the }\  U^\ell{\rm -sector}.\eqno(4.15)$$

The total physical Hilbert space $\cal H$ of strings on the ${\bf
Z}_N$-orbifold is the direct sum of the physical space $\cal H_{(\ell)}$
in the each sector:
$$ {\cal H} = {\cal H}_{(0)}\oplus{\cal H}_{(1)}\oplus \cdots \oplus{\cal
H}_{(N-1)}.\eqno(4.16)$$
In the above consideration we have shown that $\cal H$ is equivalent to
$$ {\cal H} = \{\alpha'^I_{L-n}\cdots\alpha'^J_{R-m}\cdots |
p'^I_L,p'^I_R> |n,m,\cdots \in {\bf Z} > 0,\  ({p'}^I_L,{p'}^I_R)\in {
\Gamma'}_{G}^{D,D}\},\eqno(4.17)$$
where
$$ {\Gamma'}_{G}^{D,D}=\{({p'}^I_L,{p'}^I_R)\in \bigcup_{\ell=0}^{N-1}(
\Gamma_{G}^{D,D}+\ell(v^I_L,v^I_R))\ |\ p'_L\cdot v_L - p'_R\cdot v_R -{1
\over 2}\ell ((v^I_L)^2-(v^I_R)^2) \in {\bf Z} \}.\eqno(4.18)$$
{}From eq.(4.14), ${\Gamma'}_{G}^{D,D}$ is Lorentzian even self-dual lattice
if $\Gamma_{G}^{D,D}$ is. Therefore, the total physical Hilbert space of
strings on the ${\bf Z}_N$-orbifold is nothing but that of the strings on
the torus associated with the Lorentzian even self-dual lattice ${
\Gamma'}_{G}^{D,D}$. This means that the symmetries of the ${\bf
Z}_N$-orbifold models are the same as the torus model with ${
\Gamma'}_{G}^{D,D}$.

%\endpage
\vskip 10mm

\section{5. Examples of Asymmetric ${\bf Z}_N$-Orbifolds}

{\noindent (1) Examples of Asymmetric ${\bf Z}_2$-Orbifolds}
\vskip 2mm

Let us first consider asymmetric ${\bf Z}_2$-orbifolds. The ${\bf
Z}_2$-transformation is defined by
$$ (X_L^I,X_R^I) \rightarrow (-X_L^I,X_R^I), \quad (I=1,\cdots ,D). \eqno{
(5.1)}$$
In this case, the necessary and sufficient conditions for modular
invariance (2.23), (2.24) are
$$ D=0 {\  \rm mod \  } 8 , \eqno{(5.2)}$$
$$ p_R^2 =0 {\   \rm mod \  } 1 {\qquad \rm for \  all \  } p_R \in
\Gamma_0^*, \eqno{(5.3)}$$
where
$$ \Gamma_0=\{\ p_R \ |\  (p_L=0, p_R) \in \Gamma^{D,D}\}. \eqno{(5.4)}$$
In this paper we will investigate the models with the conformal weight
(1,0) states in the twisted sector. The conformal weight of the ground
state in the twisted sector is given by $({D\over 16},0)$, so that it will
be sufficient to consider only the cases of $D=8$ and 16. For $D=16$, the
ground state in the twisted sector has the conformal weight (1,0). For $D
=8$, the ground state has the conformal weight $({1\over 2},0)$ but the
first excited state has the conformal weight (1,0) because the left-moving
oscillators are expanded in half-odd-integral modes in the twisted sector.

Let us take $\Lambda$ to be the root lattice of a simply-laced Lie group
$G$ having the ${\bf Z}_2$-automorphism shown in section 3. However,
modular invariance puts severe restrictions on $G$ or $\Lambda$ and all
the possibilities of $G$ are listed in Table 1. In the following, we will
concentrate only on the left-movers. The $G_0$ in Table 1 denotes the ${
\bf Z}_2$-invariant subgroup of $G$, which is the \lq unbroken" symmetry
in each (untwisted or twisted) left-moving Hilbert space. However, since
there appear twisted states with the conformal weight (1,0), the symmetry
will be enhanced: The physical (i.e., ${\bf Z}_2$-invariant) (1,0) states
in the untwisted sector correspond to the adjoint representation of $G_0$.
Each (1,0) state in the twisted sector corresponds to an intertwining
current which converts untwisted states to twisted ones, and vice versa.
Thus, the (1,0) states in the untwisted sector together with the (1,0)
states in the twisted sector will form an adjoint representation of a
larger group $G'$ than $G_0$, which is the full symmetry of the total
Hilbert space ($G'$ is not necessarily the same as $G$).

To investigate the enhanced symmetry $G'$, we rewrite the ${\bf
Z}_2$-orbifold model into an equivalent torus model using the
``torus-orbifold equivalence''. The equivalent torus model to the ${\bf
Z}_2$-orbifold model is specified by the following lattice:
$$ {\Gamma'}_G^{D,D}=\{(p'_L,p'_R) \in \bigcup_{\ell=0}^1 (\Gamma_G^{D,D}
+ \ell (v_L,0)) \ | \  p'_L\cdot v_L-{1\over 2}\ell v_L^2 \in {\bf Z} \},
\eqno(5.5)$$
where the shift vectors $(v_L,0)$ are
$$ \eqalign{v_L&=(1,0,0,0,0,0,0,0) \qquad {\rm for} \ G=E_8,\cr
            v_L&=({1\over2},{1\over2},{1\over2},{1\over2},0,0,0,0) \qquad
{\rm for}\  G=SO(16),\cr
            v_L&=({1\over2},{1\over2},0,0,{1\over2},{1\over2},0,0) \qquad
{\rm for}\  G=[SO(8)]^2,\cr
            v_L&=(1,0,0,0,0,0,0,0,1,0,0,0,0,0,0,0) \qquad {\rm for}\  G
=[E_8]^2, \cr
            v_L&=({1\over2},{1\over2},{1\over2},{1\over2},{1\over2},{1
\over2},{1\over2},{1\over2},0,0,0,0,0,0,0,0) \qquad {\rm for}\  G
=Spin(32)/Z_2, \cr
            v_L&=({1\over2},{1\over2},{1\over2},{1\over2},{1\over2},{1
\over2},{1\over2},{1\over2},0,0,0,0,0,0,0,0) \qquad {\rm for} \ G=SO(32),
\cr
            v_L&=({1\over2},{1\over2},{1\over2},{1\over2},{1\over2},{1
\over2},0,0,0,0,0,0,{1\over2},{1\over2},0,0) \qquad {\rm for} \ G=SO(24)
\times SO(8), \cr
            v_L&=(1,0,0,0,0,0,0,0,{1\over2},{1\over2},{1\over2},{1
\over2},0,0,0,0) \qquad {\rm for} \ G=E_8\times SO(16), \cr
            v_L&=({1\over2},{1\over2},{1\over2},{1\over2},0,0,0,0,{1
\over2},{1\over2},{1\over2},{1\over2},0,0,0,0) \qquad {\rm for} \ G=[SO
(16)]^2, \cr
            v_L&=(1,0,0,0,0,0,0,0,{1\over2},{1\over2},0,0,{1\over2},{1
\over2},0,0) \qquad {\rm for} \ G=E_8\times [SO(8)]^2, \cr
            v_L&=({1\over2},{1\over2},{1\over2},{1\over2},0,0,0,0,{1
\over2},{1\over2},0,0,{1\over2},{1\over2},0,0) \qquad {\rm for} \ G=SO(16)
\times [SO(8)]^2, \cr
            v_L&=({1\over2},{1\over2},0,0,{1\over2},{1\over2},0,0,{1
\over2},{1\over2},0,0,{1\over2},{1\over2},0,0) \qquad {\rm for} \ G=[SO
(8)]^4 \cr}$$
in the orthogonal basis. From ${\Gamma'}_G^{D,D}$, we will find what is
the enhanced symmetry $G'$. The results are summarized in Table 1.

%\endpage
\vskip 5mm
{\noindent (2) Examples of Asymmetric ${\bf Z}_3$-Orbifolds}
\vskip 2mm

Next we will discuss the asymmetric ${\bf Z}_3$-orbifolds. The ${\bf
Z}_3$-transformation is defined by
$$ (X^I_L,X^I_R) \rightarrow (U^{IJ}X^J_L,X^I_R) ,\qquad I=1,2,\cdots , D,
\eqno(5.6)$$
where $U$ is the ${\bf Z}_3$-rotation matrix, whose diagonalized matrix is
expressed by
$$ U_{diag}=diag(\omega \ \omega^2 \ \omega \ \omega^2 \ \cdots \ \omega \
\omega^2), \quad \omega=e^{2\pi i /3}.\eqno(5.8)$$

In this case, the modular invariance conditions are
$$ D=0 {\  \rm mod \  } 6 , \eqno{(5.9)}$$
$$ 3p_R^2 =0 {\   \rm mod \  } 2 {\qquad \rm for \  all \  } p_R \in
\Gamma_0^*. \eqno{(5.10)}$$
For our purpose it will be sufficient to consider only the case of $D
=6,12,18$, because the conformal weight of the ground state in the twisted
sector is $({D\over18},0)$ and a conformal weight (1,0) state in the
twisted sector appears only for $D\le 18$. For $D=18$, the ground state in
the twisted sector has the conformal weight $(1,0)$. For $D=6,12$, the
ground state in the twisted sector has the conformal weight $({6
\over18},0)$ and $({12\over18},0)$, respectively, but excited states may
have the conformal weight $(1,0)$.
Although we take $\Lambda$ having ${\bf Z}_3$-automorphism, the modular
invariance conditions restrict the allowed lattice $\Lambda$ or $G$. The
possibilities of $G$ are given in Table 2. The $G_0$ is the ${\bf
Z}_3$-invariant subgroup of $G$, which is the symmetry of each of the
untwisted and twisted Hilbert spaces. However, there exist the $(1,0)$
states in the twisted sector, so that the symmetry is enhanced and the
full symmetry of the total Hilbert space is $G'$.

To determine $G'$, we consider the equivalent torus models to the ${\bf
Z}_3$-orbifold models. The momentum lattices of the equivalent torus
models are found as follows:
$$ {\Gamma'}_G^{D,D}=\{(p'_L,p'_R) \in \bigcup_{\ell=0}^2 (\Gamma_G^{D,D}
+ \ell (v_L,0))\ | \ p'_L\cdot v_L-{1\over 2}\ell v_L^2 \in {\bf Z} \}
\eqno(5.11)$$
and the shift vectors $(v_L,0)$ are
$$\eqalign{
v_L&=v_2, \qquad{\rm for} \ G=E_6,\cr
v_L&=(v_3,v_3,v_3),\qquad{\rm for}\ G=(SU(3))^3,\cr
v_L&=(v_2,v_2),\qquad{\rm for}\ G=(E_6)^2,\cr
v_L&=(v_2,v_3,v_3,v_3),\qquad{\rm for}\ G=E_6\times(SU(3))^3,\cr
v_L&=(v_3,v_3,v_3,v_3,v_3,v_3),\qquad{\rm for}\ G=(SU(3))^6,\cr
v_L&=(v_1,v_3,v_3),\qquad{\rm for}\ G=E_8\times(SU(3))^2,\cr
v_L&=(v_2,v_2,v_2),\qquad{\rm for}\ G=(E_6)^3,\cr
v_L&=(v_2,v_2,v_3,v_3,v_3),\qquad{\rm for}\ G=(E_6)^2\times(SU(3))^3,\cr
v_L&=(v_2,v_3,v_3,v_3,v_3,v_3,v_3),\qquad{\rm for}\ G=E_6\times(SU(3))^6,
\cr
v_L&=(v_3,v_3,v_3,v_3,v_3,v_3,v_3,v_3,v_3),\qquad{\rm for}\ G=(SU(3))^9,
\cr
v_L&=(v_1,v_2,v_3,v_3),\qquad{\rm for}\ G=E_8\times E_6\times(SU(3))^2,\cr
v_L&=(v_1,v_3,v_3,v_3,v_3,v_3),\qquad{\rm for}\ G=E_8\times(SU(3))^5,\cr
v_L&=(v_1,v_1,v_3),\qquad{\rm for}\ G=(E_8)^2\times SU(3),\cr}$$
where
$$\eqalign{
v_1&\equiv{1\over3}(2\alpha_1+4\alpha_2+5\alpha_3+6\alpha_4+4\alpha_5+3
\alpha_6+3\alpha_7+\alpha_8),\cr
v_2&\equiv{1\over3}(2\beta_2+2\beta_3+3\beta_4+\beta_5+\beta_6),\cr
v_3&\equiv{1\over3}(\gamma_1+\gamma_2),\cr}$$
and $\alpha_i$, $\beta_i$ and $\gamma_i$ denote simple roots of $E_8$,
$E_6$ and $SU(3)$, respectively:

$$\eqalign{
E_8&: \qquad\matrix{
        &&        &&\alpha_2&&        &&        &&        &&      \cr
        &&        &&\circ  &&        &&        &&        &&      \cr
        &&        &&  |    &&        &&        &&        &&       \cr
\circ   &-& \circ  &-&\circ  &-& \circ  &-& \circ  &-& \circ  &-& \circ\cr
\alpha_1&&\alpha_3&&\alpha_4&&\alpha_5&&\alpha_6&&\alpha_7&&\alpha_8\cr}
\cr
&\cr
E_6&: \qquad\matrix{
        &&        &&\beta_2&&        &&        \cr
        &&        &&\circ  &&        &&        \cr
        &&        &&   |   &&        &&         \cr
\circ   &-& \circ  &-&\circ  &-& \circ  &-& \circ  \cr
\beta_1&&\beta_3&&\beta_4&&\beta_5&&\beta_6\cr}\cr
&\cr
SU(3)&: \qquad\matrix{
\circ   &-& \circ  \cr
\gamma_1&&\gamma_2\cr}\cr}$$

\vskip 5mm

\noindent
{}From this we will find what is the enhanced symmetry $G'$. The results are
summarized in Table 2.

%\endpage
\vskip 5mm
{\noindent (3) Examples of Asymmetric ${\bf Z}_N$-Orbifolds}
\vskip 2mm

In this subsection, we shall discuss asymmetric ${\bf Z}_N$-orbifolds
associated with the root lattice of $(SU(N))^n$. The root lattice of $SU
(N)$ has the ${\bf Z}_N$-symmetry, i.e., the cyclic permutation symmetry:
$\alpha_i \rightarrow \alpha_{i+1}$ $(i=0,1,\cdots ,N-1)$ where $\alpha_i$
$(i=1,\cdots ,N-1)$ is a simple root of $SU(N)$ and $\alpha_0=\alpha_N=-(
\alpha_1+\alpha_2+\cdots +\alpha_{N-1})$. Then, we define the ${\bf
Z}_N$-transformation by
$$ g: \quad (X_L^I,X_R^I) \rightarrow (U^{IJ}X_L^J,X_R^I), \eqno{(5.12)}$$
where $U$ is the $n(N-1)\times n(N-1)$ rotation matrix which generates the
above ${\bf Z}_N$-cyclic permutation and its diagonalized matrix is
$$ U_{diag}=diag(\omega \ \omega^2 \ \cdots \ \omega^{N-1} \ \omega \
\omega^2 \ \cdots \ \omega^{N-1} \ \cdots \ \omega \ \omega^2 \ \cdots \
\omega^{N-1}), \quad \omega=e^{2\pi i / N}.\eqno(5.13)$$

The conditions for modular invariance and the existence of (1,0) twisted
states put severe restrictions on the allowed values of $N$ and $n$. All
the models we have to consider are shown in Table 3. The $G_0$ is the ${
\bf Z}_N$-invariant subgroup of $G$, which is the symmetry in each of the
untwisted and twisted Hilbert spaces. The full symmetry of the total
Hilbert space is denoted by $G'$.

To determine $G'$, we consider the equivalent torus model to the ${\bf
Z}_N$-orbifold model. The momentum lattice of the equivalent torus model
is found as follows:
$$ {\Gamma'}_{(SU(N))^n}^{D,D}=\{(p'_L,p'_R) \in \bigcup_{\ell=0}^{N-1} (
\Gamma_{(SU(N))^n}^{D,D} + \ell (v_L,0)) \ |  \  p'_L\cdot v_L-{1\over 2}
\ell v_L^2 \in {\bf Z} \},\eqno(5.14)$$
where the shift vector $(v_L,0)$ is
$$v_L=(v_L^{(1)},v_L^{(2)},\cdots,v_L^{(n)}),$$
$$v_L^{(i)}={1\over 2N}\sum_{j=1}^{N-1}j(N-j)\alpha_j \qquad {\rm for \
all} \ i.$$
Thus we will find what is  the enhanced symmetry $G'$. The results are
summarized in Table 3.

%\endpage
\vskip 10mm

\section{6. Conclusion}
We have shown various examples of ${\bf Z}_N$-asymmetric orbifold models
to possess twist-untwist intertwining currents and investigated the
symmetries of them. Since every physical string state must be invariant
under the ${\bf Z}_N$-transformation, the ${\bf Z}_N$-invariant subgroup
$G_0$ is the ``unbroken'' symmetry in each of the untwisted and twisted
sectors. However, when there exist intertwining currents, the currents
convert untwisted states to twisted ones and vice versa. Therefore the
symmetry is enhanced to a larger group $G'$ than $G_0$. We have seen that
the conditions for the lattice with ${\bf Z}_N$-automorphism, modular
invariance and the existence of (1,0) twisted states put severe
restrictions on such orbifold models.

We have here investigated ${\bf Z}_2$- and ${\bf Z}_3$-orbifold models
associated with the root lattices of simply-laced Lie groups and ${\bf
Z}_N$-orbifold models with the root lattices of $(SU(N))^n$, where the ${
\bf Z}_N$-transformation leaves only the origin fixed. The remaining task
is to find all other ${\bf Z}_N$-orbifold models which possess
intertwining currents. The complete list of such orbifold models will be
reported elsewhere.

In this paper, we have discussed the case that rank of $G$ is equal to
the dimension of the orbifold, i.e., the case that the ${\bf
Z}_N$-transformation is an inner automorphism, and we have investigated
the symmetry of the orbifold model by rewriting it into an equivalent
torus model. In the previous paper [9], we have also found the asymmetric
orbifold models which can probably be rewritten into torus models though
the orbifolds are defined by outer automorphisms of the momentum lattices.

It will be straightforward to apply our analysis to the heterotic string
theory [22]. We hope to get new phenomenologically interesting superstring
models along this line.

\endpage
\vskip 10mm

\references
\item{[1]} L. Dixon, J.A. Harvey, C. Vafa and E. Witten, Nucl. Phys. {\bf
B261} (1985) 678; {\bf B274} (1986) 285.
\item{[2]} A. Font, L.E. Ib\'a\~ nez, F. Quevedo and A. Sierra, Nucl.
Phys. {\bf B331} (1990) 421.
\item{[3]} J.A. Casas and C. Mu\~ noz, Nucl. Phys. {\bf B332} (1990) 189.
\item{[4]} Y. Katsuki, Y. Kawamura, T. Kobayashi, N. Ohtsubo, Y. Ono and
K. Tanioka, Nucl. Phys. {\bf B341} (1990) 611.
\item{[5]} A. Fujitsu, T. Kitazoe, M. Tabuse and H. Nishimura, Int. J.
Mod. Phys. {\bf A5} (1990) 1529.
\item{[6]} K.S. Narain, M.H. Sarmadi and C. Vafa, Nucl. Phys. {\bf B288}
(1987) 551; {\bf B356} (1991) 163.
\item{[7]} E. Corrigan and T.J. Hollowood, Nucl. Phys. {\bf B304} (1988)
77.
\item{[8]} L. Dolan, P. Goddard and P. Montague, Nucl. Phys. {\bf B338}
(1990) 529.
\item{[9]} Y. Imamura, M. Sakamoto and M. Tabuse, Phys. Lett. {\bf B266}
(1991) 307.
\item{[10]} E. Corrigan and D. Olive, Nuovo Cimento {\bf 11A} (1972) 749.
\item{[11]} Y. Kazama\quad  and\quad  H. Suzuki,\quad  Phys. Lett. {\bf
B192} (1987) 351;\quad  KEK preprint KEK-TH-165 (1987).
\item{[12]} B. Gato, Nucl. Phys. {\bf B322} (1989) 555.
\item{[13]} I. Frenkel, J. Lepowsky and A. Meurman, Proc. Conf. on Vertex
operators in mathematics and physics, ed. J. Lepowsky et al. (Springer
1984).
\item{[14]} R. Dijkgraaf, E. Verlinde and H. Verlinde, Commun. Math. Phys.
{\bf 115} (1988) 649;
\item{    } P. Ginsparg, Nucl. Phys. {\bf B295} (1988) 153.
\item{[15]} M. Sakamoto, Mod. Phys. Lett. {\bf A5} (1990) 1131; Prog.
Theor. Phys. {\bf 84} (1990) 351.
\item{[16]} K.S. Narain, Phys. Lett. {\bf B169} (1986) 41;
\item{    } K.S. Narain, M.H. Sarmadi and E. Witten, Nucl. Phys. {\bf
B279} (1987) 369.
\item{[17]} C. Vafa, Nucl. Phys. {\bf B273} (1986) 592.
\item{[18]} J.E. Humphreys, \ Introduction \ to \ Lie \ Algebras \ and \
Representation \ Theory \  (Springer 1972);
\item{    } N. Bourbaki, {\sl \'El\'ements  de  Math\'ematique, Groupes et
Algebras  Lie }  (Paris: Hermann 1960).
\item{[19]} R.G. Myhill, Univ. of Durham preprint DTP-86/19 (1986).
\item{[20]} W. Lerche, A.N. Schellekens and N.P. Warner, Phys. Rep. {\bf
177} (1989) 1.
\item{[21]} I. Frenkel and V. Kac, Inv. Math. {\bf 62} (1980) 23;
\item{    } G. Segal, Commun. Math. Phys. {\bf 80} (1981) 301.
\item{[22]} D.J. Gross, J.A. Harvey, E. Martinec and R. Rohm, Nucl. Phys.
{\bf B256} (1985) 253; {\bf B267} (1986) 75.

\endpage
%\vskip 10mm

\table
\noindent
Table 1. Examples of asymmetric ${\bf Z}_2$-orbifold models: $G_0$ denotes
the ${\bf Z}_2$-invariant subgroup of $G$ which is the symmetry in each
sector and $G'$ denotes the full symmetry of the total Hilbert space.

\vskip 2mm

\noindent
Table 2. Examples of asymmetric ${\bf Z}_3$-orbifold models: $G_0$ denotes
the ${\bf Z}_3$-invariant subgroup of $G$ which is the symmetry in each
sector and $G'$ denotes the full symmetry of the total Hilbert space.

\vskip 2mm

\noindent
Table 3. Examples of asymmetric ${\bf Z}_N$-orbifold models: $G_0$ denotes
the ${\bf Z}_N$-invariant subgroup of $G$ which is the symmetry in each
sector and $G'$ denotes the full symmetry of the total Hilbert space.

\endpage

\nopagenumbers

%%%%%%%%%%% Table 1. %%%%%%%%%%%%%%%%

\centerline{
\vbox{\offinterlineskip
\halign{&\vrule#&\strut\quad\hfil#\hfil\quad&\vrule#&\quad\hfil#\hfil
\quad&\vrule#&\quad\hfil#\hfil\quad&\vrule# \cr
\noalign{\hrule}
height3pt&\omit      &&\omit             &&\omit            &\cr
&  $G$                && $G_0$            && $G'$             &\cr
height3pt&\omit      &&\omit             &&\omit            &\cr
\noalign{\hrule}
height4pt&\omit      &&\omit             &&\omit            &\cr
& $D=8$\qquad\qquad\qquad\quad &&                &&                &\cr
height3pt&\omit      &&\omit             &&\omit            &\cr
& $E_8$              && $SO(16)$         && $E_8$            &\cr
height3pt&\omit      &&\omit             &&\omit            &\cr
& $SO(16)$           && $(SO(8))^2$      && $SO(16)$         &\cr
height3pt&\omit      &&\omit             &&\omit            &\cr
& $(SO(8))^2$        && $(SU(2))^8$      && $(SO(8))^2$      &\cr
height4pt&\omit      &&\omit             &&\omit            &\cr
& $D=16$\qquad\qquad\qquad\  &&                &&                &\cr
height3pt&\omit      &&\omit             &&\omit            &\cr
& $(E_8)^2$          && $(SO(16))^2$     && $Spin(32)/Z_2$   &\cr
height3pt&\omit      &&\omit             &&\omit            &\cr
& $Spin(32)/Z_2$     && $(SO(16))^2$     && $(E_8)^2$        &\cr
height3pt&\omit      &&\omit             &&\omit            &\cr
& $SO(32)$           && $(SO(16))^2$     && $E_8\times SO(16)$ &\cr
height3pt&\omit      &&\omit             &&\omit            &\cr
& $SO(24)\times SO(8)$ && $(SO(12))^2\times (SU(2))^4$ && $E_7\times SO
(12)\times (SU(2))^3$ &\cr
height3pt&\omit      &&\omit             &&\omit            &\cr
& $E_8\times SO(16)$ && $SO(16)\times (SO(8))^2$ && $SO(24)\times SO(8)$ &
\cr
height3pt&\omit      &&\omit             &&\omit            &\cr
& $(SO(16))^2$ && $(SO(8))^4$ && $SO(16)\times (SO(8))^2$ &\cr
height3pt&\omit      &&\omit             &&\omit            &\cr
& $E_8\times (SO(8))^2$ && $SO(16)\times (SU(2))^8$ && $SO(20)\times (SU
(2))^6$ &\cr
height3pt&\omit      &&\omit             &&\omit            &\cr
& $SO(16)\times (SO(8))^2$ && $(SO(8))^2\times (SU(2))^8$ && $SO(12)\times
SO(8)\times (SU(2))^6$ &\cr
height3pt&\omit      &&\omit             &&\omit            &\cr
& $(SO(8))^4$ && $(SU(2))^{16}$ && $SO(8)\times (SU(2))^{12}$ &\cr
height4pt&\omit      &&\omit             &&\omit            &\cr
\noalign{\hrule}
}}}
%\vskip 4mm
\centerline{\bf Table 1}

\endpage

%%%%%%%%%%%%%%% Table 2 %%%%%%%%%%%%%%%%%

\centerline{
\vbox{\offinterlineskip
\halign{&\vrule#&\strut\quad\hfil#\hfil\quad&\vrule#&\quad\hfil#\hfil
\quad&\vrule#&\quad\hfil#\hfil\quad&\vrule# \cr
\noalign{\hrule}
height3pt&\omit      &&\omit             &&\omit            &\cr
&  $G$                && $G_0$            && $G'$             &\cr
height3pt&\omit      &&\omit             &&\omit            &\cr
\noalign{\hrule}
height4pt&\omit      &&\omit             &&\omit            &\cr
& $D=6$\qquad\qquad\qquad\quad &&                &&                &\cr
height3pt&\omit      &&\omit             &&\omit            &\cr
& $E_6$              && $(SU(3))^3$      && $E_6$            &\cr
height3pt&\omit      &&\omit             &&\omit            &\cr
& $(SU(3))^3$        && $(U(1))^6$       && $(SU(3))^3$     &\cr
height4pt&\omit      &&\omit             &&\omit            &\cr
& $D=12$\qquad\qquad\qquad\  &&                &&                &\cr
height3pt&\omit      &&\omit             &&\omit            &\cr
& $(E_6)^2$        && $(SU(3))^6$      && $(E_6)^2$      &\cr
height3pt&\omit      &&\omit             &&\omit            &\cr
& $E_6\times(SU(3))^3$ && $(SU(3))^3\times(U(1))^6$ && $SU(6)\times SO(8)
\times(U(1))^3$      &\cr
height3pt&\omit      &&\omit             &&\omit            &\cr
& $(SU(3))^6$        && $(U(1))^{12}$      && $(SU(2))^6\times(U(1))^6$
&\cr
height3pt&\omit      &&\omit             &&\omit            &\cr
& $E_8\times(SU(3))^2$ && $SU(9)\times(U(1))^4$ && $SO(20)\times(U(1))^2$
&\cr
height4pt&\omit      &&\omit             &&\omit            &\cr
& $D=18$\qquad\qquad\qquad\  &&                &&                &\cr
height3pt&\omit      &&\omit             &&\omit            &\cr
& $(E_6)^3$          && $(SU(3))^9$     && $E_6\times(SU(3))^6$   &\cr
height3pt&\omit      &&\omit             &&\omit            &\cr
& $(E_6)^2\times(SU(3))^3$ && $(SU(3))^6\times(U(1))^6$ &&$SU(6)\times(SU
(3))^4\times(U(1))^5$        &\cr
height3pt&\omit      &&\omit             &&\omit            &\cr
& $E_6\times(SU(3))^6$ && $(SU(3))^3\times(U(1))^{12}$ && $SU(4)\times(SU
(3))^2\times(U(1))^{11}$ &\cr
height3pt&\omit      &&\omit             &&\omit            &\cr
& $(SU(3))^9$ && $(U(1))^{18}$ && $SU(2)\times(U(1))^{17}$ &\cr
height3pt&\omit      &&\omit             &&\omit            &\cr
& $E_8\times E_6\times(SU(3))^2$ && $SU(9)\times(SU(3))^3\times(U(1))^4$
&& $SU(12)\times(SU(3))^2\times(U(1))^3$ &\cr
height3pt&\omit      &&\omit             &&\omit            &\cr
& $E_8\times(SU(3))^5$ && $SU(9)\times(U(1))^{10}$ && $SU(10)\times(U
(1))^{9}$ &\cr
height3pt&\omit      &&\omit             &&\omit            &\cr
& $(E_8)^2\times SU(3)$ && $(SU(9))^2\times(U(1))^2$ && $SU(18)\times U
(1)$ &\cr
height4pt&\omit      &&\omit             &&\omit            &\cr
\noalign{\hrule}
}}}
%\vskip 4mm
\centerline{\bf Table 2}

\endpage

%%%%%%%%%%%%% Table 3 %%%%%%%%%%%%%%%%%

\centerline{
\vbox{\offinterlineskip
\halign{&\vrule#&\strut\quad\hfil#\hfil\quad&\vrule#&\quad\hfil#\hfil
\quad&\vrule#&\quad\hfil#\hfil\quad&\vrule# \cr
\noalign{\hrule}
height3pt&\omit      &&\omit             &&\omit            &\cr
&  $G$                && $G_0$            && $G'$             &\cr
height3pt&\omit      &&\omit             &&\omit            &\cr
\noalign{\hrule}
height3pt&\omit      &&\omit             &&\omit            &\cr
& $(SU(3))^3$  && $(U(1))^6$         && $(SU(3))^3$   &\cr
height3pt&\omit      &&\omit             &&\omit            &\cr
& $(SU(3))^6$ && $(U(1))^{12}$         && $(SU(2))^6\times (U(1))^6$   &
\cr
height3pt&\omit      &&\omit             &&\omit            &\cr
& $(SU(3))^9$ && $(U(1))^{18}$         && $SU(2)\times (U(1))^{17}$   &\cr
height3pt&\omit      &&\omit             &&\omit            &\cr
& $SU(5)$ && $(U(1))^4$         && $SU(5)$   &\cr
height3pt&\omit      &&\omit             &&\omit            &\cr
& $(SU(5))^2$ && $(U(1))^8$         && $(SU(5))^2$   &\cr
height3pt&\omit      &&\omit             &&\omit            &\cr
& $(SU(5))^3$ && $(U(1))^{12}$         && $(SU(4))^3\times (U(1))^3$   &
\cr
height3pt&\omit      &&\omit             &&\omit            &\cr
& $(SU(5))^4$ && $(U(1))^{16}$         && $(SU(2))^8\times (U(1))^8$   &
\cr
height3pt&\omit      &&\omit             &&\omit            &\cr
& $(SU(5))^5$ && $(U(1))^{20}$         && $(SU(2))^2\times (U(1))^{18}$
&\cr
height3pt&\omit      &&\omit             &&\omit            &\cr
& $SU(7)$ && $(U(1))^6$         && $SU(7)$   &\cr
height3pt&\omit      &&\omit             &&\omit            &\cr
& $(SU(7))^2$ && $(U(1))^{12}$         && $(SU(6))^2\times (U(1))^2$   &
\cr
height3pt&\omit      &&\omit             &&\omit            &\cr
& $(SU(7))^3$ && $(U(1))^{18}$         && $(SU(2))^9\times (U(1))^9$   &
\cr
height3pt&\omit      &&\omit             &&\omit            &\cr
& $SU(11)$ && $(U(1))^{10}$         && $SU(11)$   &\cr
height3pt&\omit      &&\omit             &&\omit            &\cr
& $(SU(11))^2$ && $(U(1))^{20}$      && $(SU(2))^{10}\times (U(1))^{10}$
&\cr
height3pt&\omit      &&\omit             &&\omit            &\cr
& $SU(13)$ && $(U(1))^{12}$         && $SU(12)\times U(1)$   &\cr
height3pt&\omit      &&\omit             &&\omit            &\cr
& $SU(17)$ && $(U(1))^{16}$         && $(SU(8))^2\times (U(1))^2$   &\cr
height3pt&\omit      &&\omit             &&\omit            &\cr
& $SU(19)$      && $(U(1))^{18}$         && $(SU(6))^3\times (U(1))^3$   &
\cr
height3pt&\omit      &&\omit             &&\omit            &\cr
& $SU(23)$      && $(U(1))^{22}$     && $(SU(2))^{11}\times (U(1))^{11}$
&\cr
height3pt&\omit      &&\omit             &&\omit            &\cr
\noalign{\hrule}
}}}
\vskip 4mm
\centerline{\bf Table 3}

\end